\documentclass[twocolumn,showpacs,preprintnumbers,amsmath,amssymb, prl]{revtex4}

\usepackage{graphicx}% Include figure files
\usepackage{dcolumn}% Align table columns on decimal point
\usepackage{bm}% bold math
\usepackage{color}

\begin{document}

\preprint{}

\title{Local control of spin flow in weak magnetic fields}

\author{J.-H.~Quast}
\author{G.~V.~Astakhov}
\altaffiliation{Also at A.~F.~Ioffe Physico-Technical Institute,
RAS, 194021 St. Petersburg, Russia} \email[\\ E-mail:
]{astakhov@physik.uni-wuerzburg.de}
\author{W.~Ossau}
\author{L.~W.~Molenkamp}
\affiliation{Physikalisches Institut (EP3), Universit\"{a}t
W\"{u}rzburg, 97074 W\"{u}rzburg, Germany}

\author{J.~Heinrich}
\author{S.~H\"{o}fling}
\author{A.~Forchel}
\affiliation{Physikalisches Institut (TP), Universit\"{a}t
W\"{u}rzburg, 97074 W\"{u}rzburg, Germany}

\date{\today}

\begin{abstract}
We demonstrate that optical illumination strongly influences spin
transport in n-type GaAs. Specifically, increasing the power
density of optical spin pumping results in a significant expansion
of the spin diffusion profile. A further means of manipulation is
the application of a weak transverse magnetic field, which
strongly increases spin flow out of the excitation spot. These
effects are directly monitored in spin imaging experiments and
spatially resolved Hanle measurements.
\end{abstract}


\maketitle
%---------------------------------------------------------------

%\textcolor{red}{In case of major corrections or comments use
%color...}

The ability to monitor, control and manipulate spin flows in
semiconductors is a prerequisite for the functionality of
spin-based devices. Because of the spin-selectivity of the
selection rules \cite{OO}, optical spectroscopy has emerged as a
powerful tool to locally probe the spin of electrons. Spin
diffusion lengths of over ca. $10$~$\mathrm{\mu m}$ in n-GaAs
\cite{Dzh1} and spin drag over $100$~$\mathrm{\mu m}$
\cite{SpinFlow1} enable optical detection of spin injection
\cite{SpinDrift1,SpinDrift2,SpinDrift3, SpinFlow2, SpinInjCross},
spin accumulation \cite{SpinFlow2} and the spin Hall effect
\cite{SHE}. In this type of experiments, the optical excitation
density is kept low in order to minimize a possible perturbation
of the spin system, implicitly assuming that using a low optical
power density implies that the dilution of the intrinsic electrons
in the semiconductor with photo-generated ones can be neglected.
Generally, this condition is met when $G \tau < n_e$, where $G$ is
the generation rate (which is proportional to the illumination
power density), $\tau$ is the lifetime of the photo-generated
carriers and $n_e$ is the concentration of intrinsic electrons.
However, for a spin polarized electron system, recombination with
photo-generated holes will reduce the spin polarization. This
additional effect becomes relevant when $G \tau_s > n_e$, where
$\tau_s$ now is the electron spin relaxation time \cite{OO}.
Therefore, in semiconductors with a very long electron spin
memory, i.~e., when $\tau_s \gg \tau$, this mechanism may lead to
enhanced spin decay at quite low pump/probe power densities, and
hence should not be neglected a priori. On the plus side, this
same mechanism can be used to locally control spin flow even
without application of a bias voltage, as we will demonstrate
below.
% Laser light creates a spatial profile of spin
% lifetimes ($T_s$), and thus the suppression of spin polarization
% in a transverse magnetic field (known as the Hanle effect
% \cite{OO}) becomes spatially dependent. In particular, stronger
% spin pumping results in a shorter spin lifetime and thus a
% stronger magnetic field is required to suppress the spin
% polarization within the pump spot. While outside the pump spot the
% spin polarization reduces more quickly with rising fields. As a
% result, in weak magnetic fields (a few tens of Gauss) the spin
% gradient (and therefore the spin outflow from the injection area)
% increases.

We present results for a Si-doped  n-type GaAs layer ($n_e = 2.5
\times 10^{16}$~$\mathrm{cm^{-3}}$) of $d = 1.5$~$\mathrm{\mu m}$
width. It was grown by molecular-beam epitaxy (MBE) on
semi-insulating (001) GaAs substrate followed by an undoped 200~nm
GaAs buffer, a 5~nm AlAs barrier and an undoped 100~nm GaAs spacer
layer. The sample is mounted strain-free and kept at a temperature
$T = 8$~K. In order to perform optical spin pumping and probing we
use a two-color Kerr rotation technique \cite{Method_TwoColor}.
Optical excitation is performed by a solid state laser (785~nm)
modulated between $\sigma^+$ and $\sigma^-$ circular polarizations
at a frequency of 50~kHz. The net spin polarization along $z$
direction $S_z$ is probed using the magneto-optical Kerr effect
(MOKE). The photoinduced Kerr rotation $\theta$ of a Ti:sapphire
laser (819~nm) which is proportional to the spin polarization
($\theta \propto S_z$) is measured by balanced photodiodes and
demodulated by a lock-in amplifier. Scanning Kerr microscopy is
used to spatially resolve the net spin polarization
\cite{SpinFlow1,Method_ScanKerr}. In this technique, a
circularly-polarized pump beam is directed under a $45^{\circ}$
angle of incidence to the sample surface. After refraction inside
the sample the pump beam generates spins polarized at an angle
$\gamma = 11^{\circ}$ with respect to the sample normal ($z$
direction). Surface scans in the $x$-$y$ plane are performed using
a microscope objective (NA = 0.14) mounted on a piezo system.
Another microscope objective is used to focus the pump beam.

\begin{figure}[tb]
\includegraphics[width=.43\textwidth]{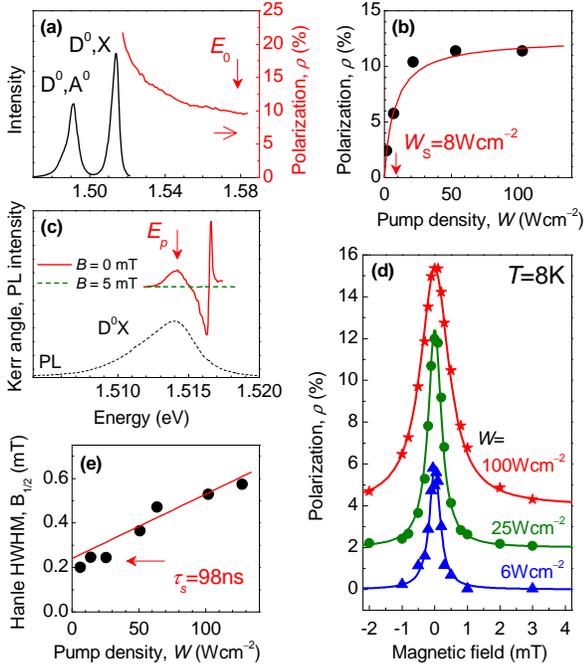}
\caption{(Color online) Sample characterization in a configuration
where spin diffusion is averaged out. (a) PL and polarization
excitation spectra. (b) PL polarization vs. pump power density.
Solid line is a fit to Eq.~(\ref{Eq0}). (c) Kerr angle $\theta$
vs. probe energy $E_p$. Pump energy is $E_0 = 1.58$~eV. (d) Hanle
curves for different pump power densities. Solid lines represent
fits to Eq.~(\ref{Eq_Hanle}). The curves are shifted for clarity.
(e) Hanle HWHM $B_{1/2}$ vs. pump power density $W$. The solid
line is a fit to Eq.~(\ref{Eq1}). } \label{fig_NonDiff}
\end{figure}

First, we characterize the sample in an excitation configuration
where spin diffusion is averaged out (Fig.~\ref{fig_NonDiff}),
using two-color Kerr and photoluminescence (PL) measurements.
Instead of the microscope objective, we employ two long-focus
lenses to provide spot sizes of about $100$~$\mathrm{\mu m}$, much
larger than the typical spin diffusion length. The PL spectrum
consists of two bands which we attribute to donor-acceptor
($\mathrm{D^{0},A^{0}}$) transitions and donor-bound excitonic
($\mathrm{D^{0},X}$) recombination [see
Fig.~\ref{fig_NonDiff}(a)]. In the same Figure, we plot with the
optically induced circular polarization $\rho_c$ of the PL
[detected at ($\mathrm{D^{0},X}$) band], as function of the
excitation energy. A maximum value of $\rho_c = 22\%$ (the
theoretical limit is 25\% \cite{OO}) is achieved for
quasi-resonant excitation. The vertical arrow in
Fig.~\ref{fig_NonDiff}(a) indicates the pump energy $E_0 =
1.58$~eV which will be used in all further experiments. The PL
circular polarization is proportional to the net spin
polarization, $\rho_c=S_z / 2$. Its amplitude depends on pump
power;  $\rho$ increases linearly for low- and saturates for high
pump power density, as shown in Fig.~\ref{fig_NonDiff}(b). This
behavior can be well described  by \cite{OO}
\begin{equation}
S_z (W) = \frac{S_{z0}}{1+ W_S / W}   \label{Eq0}
\end{equation}
using $S_{z0} = 25 \%$ (i.e., $\rho_{c0} = 12.5 \%$) and a
saturation pump power density $W_S = 8$~$\mathrm{W cm^{-2}}$. The net
spin polarization is suppressed in a transverse magnetic field
$B$ due to the Hanle effect \cite{OO}. The Hanle curves we obtained for this sample are
shown in Fig.~\ref{fig_NonDiff}(d). They are well described by the
Lorentzian
\begin{equation}
S_z = \frac{S_z (W)}{1 + (B / B_{1/2})^2} \,.   \label{Eq_Hanle}
\end{equation}
Here, $B_{1/2} = \hbar / (g_e \mu_B) T_s^{-1}$ is the halfwidth at
half maximum (HWHM) of the Hanle curve given by the spin lifetime
$T_s$ connected with optical recombination. The Hanle HWHM $B_{1/2}$ increases with pump power density $W$
as shown in Fig.~\ref{fig_NonDiff}(e). This behavior is well
established for n-GaAs \cite{OO}. Upon increasing $W$, recombination
with photogenerated holes provides an additional spin decay
channel, resulting in a decrease of the spin lifetime given by
\cite{Method_TwoColor}
\begin{equation}
B_{1/2} \propto T_s^{-1} = \tau_s^{-1} (1 + W / W_0) \,.
\label{Eq1}
\end{equation}
Here, $W_0$ defines a characteristic pump power density, and the
regime where  $W > W_0$ implies strong spin pumping. In the low
power density limit, $T_s$ is equal to the spin relaxation time of
electrons $\tau_s$, and, using the well-known electron g-factor in
GaAs ($g_e = -0.44$) our Hanle data yield an electron spin
lifetime  $\tau_s = 98$~ns [Fig.~\ref{fig_NonDiff}(e)].

\begin{figure}[b]
\includegraphics[width=.43\textwidth]{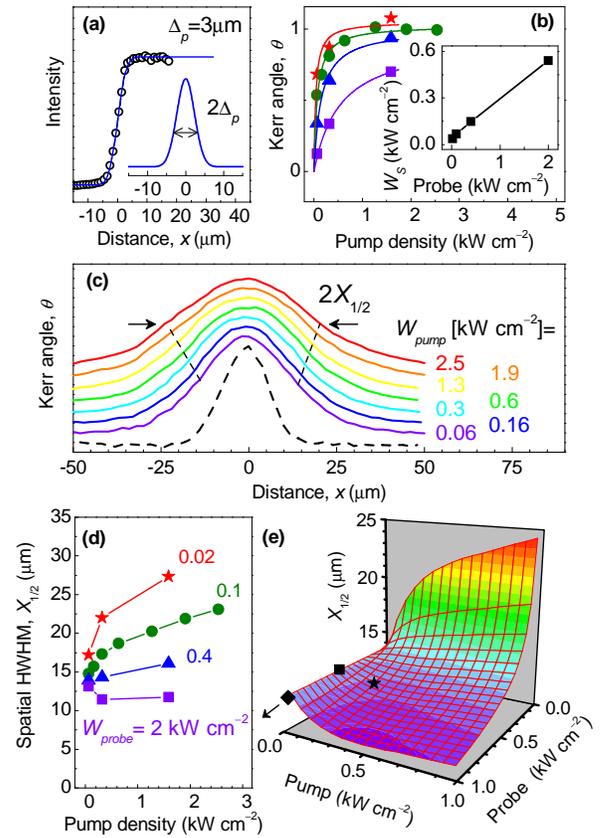}
\caption{(Color online) Spin pumping in the diffusive regime. (a) Scan
of the laser spot through a sharp edge. Inset: laser spot profile.
(b) Kerr angle at the center of the pump spot ($x = 0$) vs. pump
power density for several probe power densities. Solid lines are fits to
Eq.~(\ref{Eq0}). Inset: saturation pump power density $W_S$ vs. probe
power density $W_{probe}$. (c) Spatial spin distribution for different
pump power densities $W_{pump}$. The probe power density is $W_{probe} =
0.1$~$\mathrm{kW cm^{-2}}$. The dotted curve gives the experimental
resolution. (d) Spatial HWHM $X_{1/2}$ vs. pump power density for
several probe power densities. (e) Overview of the 2D dependence of the
spatial HWHM $X_{1/2}$ on pump and probe power density. Symbols
indicate experimental conditions in various papers: $\bigstar$ -
Ref.~\onlinecite{SpinFlow1}, $\blacksquare$ -
Ref.~\onlinecite{SpinFlow2} and $\blacklozenge$ -
Ref.~\onlinecite{SHE}. } \label{fig_DiffPump}
\end{figure}

The Hanle effect is also clearly observed in MOKE measurements.
Figure~\ref{fig_NonDiff}(c) demonstrates the dependence of the
Kerr angle $\theta$ on the probe energy $E_p$ in zero magnetic
field (solid line). Note that the Kerr angle completely vanishes at $B = 5$~mT (in
the limit of low pump and probe density, dotted line). In the subsequent experiments,
we measure the Kerr angle at a fixed energy of $E_p = 1.514$~eV (as indicated by the arrow).

%%%%%%%%%%%%%%%%%%%%%%%%%%%%%%%%%%%%%%%%%%%%%%%%%%%%%%%%

We now describe our results obtained by Kerr microscopy, where the
laser beams are focused tight enough to resolve spin diffusion.
The spatial resolution of our setup can be inferred from a scan of
the transmission of the focused laser beam over a sharp edge, as
shown in Fig.~\ref{fig_DiffPump}(a). The derivative of this trace
(inset) confirms that the spot profile has the form $\exp(-x^2 /
\Delta_p^2)$, with $\Delta_p = 3.3$~$\mathrm{\mu m}$; we have
verified that the profile in the sharp-edge scans is circular.
Diffusion of optically injected spins causes the spatial profiles
shown in Fig.~\ref{fig_DiffPump}(c). The dotted line in this
figure shows the excitation profile as detected by the probe. It
corresponds to the convolution of pump and probe spots and yields
a net resolution of $\Delta = 7.5$~$\mathrm{\mu m}$. The spatial
HWHM $X_{1/2}$ of the scans in the Figure (and therefore the spin
diffusion length $L_s = \sqrt{D_s \tau_s}$) is not constant, as
one might naively anticipate, but rises from 14~$\mathrm{\mu m}$
for $W = 0.06$~$\mathrm{kW cm^{-2}}$ to 23~$\mathrm{\mu m}$ for $W
= 2.5$~$\mathrm{kW cm^{-2}}$. It seems unlikely that the spin
relaxation time $\tau_s$ becomes longer with increasing spin
pumping. Hence, we ascribe such a behavior to an increase of the
spin diffusion constant $D_s$. This suggests that optically
injected electrons not only provide a local spin source, but also
participate in spin transfer over tens of microns. Note that a
possible heating effect by light can be ruled out, as we observe a
decrease of $X_{1/2}$ with rising bath temperature.

This effect is summarized for different probe power densities in
Fig.~\ref{fig_DiffPump}(d). As a general trend, the spin diffusion
length decreases with increasing probe power density. In order to
examine the origin of such a behavior, we plot in
Fig.~\ref{fig_DiffPump}(b) the Kerr rotation at the excitation
point ($x = 0$) as function of pump power density for several
probe power densities. These dependencies are well fitted by
Eq.~(\ref{Eq0}), where the fitting procedure reveals that also the
saturation pump power density $W_S$ depends on the probe power
density, following a linear increase [see inset of
Fig.~\ref{fig_DiffPump}(b)]. This makes intuitive sense: the
faster the spin decay, the stronger spin pumping is required to
achieve saturation. The probe beam generates holes, introducing an
additional spin decay channel. The efficiency of this channel
scales with the ratio of the spin relaxation time and the
electron-hole recombination time, and it obviously is efficient
even at moderate power density. As a consequence, the spin
polarization is destroyed and the spin diffusion is suppressed for
increasing probe power density. The overall dependence of the spin
diffusion process on the pump and probe power density is
summarized in Fig.~\ref{fig_DiffPump}(e). In this plot we have
included the actual experimental conditions from various recent
papers, labeled by symbols. While the influence of the power level
used in these works is not very significant, these data have not
been taken in the true low perturbation regime.

\begin{figure}[tbp]
\includegraphics[width=.45\textwidth]{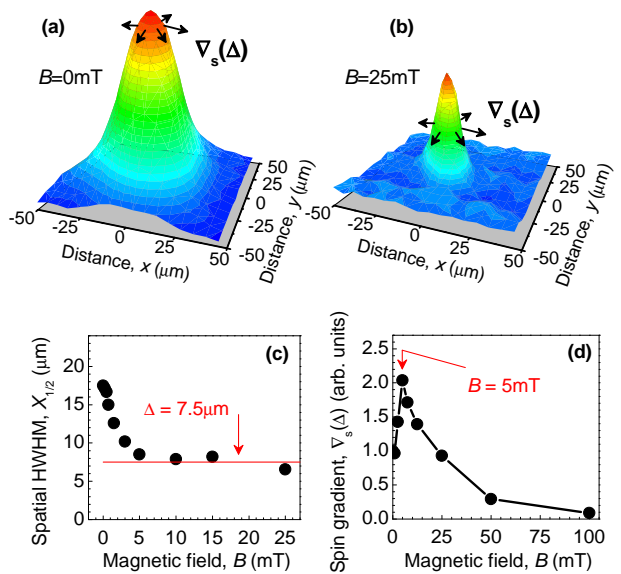}
\caption{(Color online) (a) Spatial spin distribution in zero
magnetic field. (b) The same in a transverse magnetic field of
25~mT. (c) Spatial HWHM vs. magnetic field. (d) Spin gradient
$\partial \theta /
\partial x \propto \nabla_s$ obtained at a distance $x = \Delta$
vs. magnetic field. $W_{pump} = 1.6$~$\mathrm{kW cm^{-2}}$ and
$W_{probe} = 0.1$~$\mathrm{kW cm^{-2}}$. } \label{fig_2D}
\end{figure}

We now demonstrate how the above effect in combination with weak magnetic
fields can be used to locally manipulate the spin flow around the excitation spot.
A 2D spatial scan in the regime of strong spin pumping
$W_{pump} \gg W_{0}$ ($W_{pump} = 1.6$~$\mathrm{kW cm^{-2}}$)
is shown in Fig.~\ref{fig_2D}(a) for zero magnetic field. In order to
minimize the influence of the probe light, its power density is reduced
to $W_{probe} = 0.1$~$\mathrm{kW cm^{-2}}$. When an external
magnetic field is applied in the sample plane, the spatial spin
distribution changes drastically [Fig.~\ref{fig_2D}(b)]. As
clearly seen in Fig.~\ref{fig_2D}(c), the HWHM  of the spatial
profile $X_{1/2}$ decreases with increasing magnetic field and
saturates for $B > 5$~mT at the resolution limit of $\Delta = 7.5$~$\mathrm{\mu
m}$ resulting from the finite sizes of pump and probe spots.

Obviously, also the spin gradient $\nabla_s \propto \partial \theta /
\partial x $ in Fig.~\ref{fig_2D}
depends on the in-plane magnetic field. We now concentrate on the spin
gradient at the point $x = \Delta$. It has a precise physical
meaning: $\nabla_s (\Delta)$ is proportional to the spin flow emanating
from the injection point due to the diffusion process. Hence, we
plot in Fig.~\ref{fig_2D}(d) $\partial \theta /
\partial x$ obtained at $x = \Delta$ as function of the
magnetic field. The data are normalized by their value in zero
field. The spin gradient shows a non-monotonic behavior. For small
magnetic fields it increases until its value has doubled, then it
decreases towards zero in stronger fields. The maximum is achieved
at $B = 5$~mT. This implies that the spin flow from the
injection area is enhanced in weak in-plane magnetic fields.

\begin{figure}[tb]
\includegraphics[width=.41\textwidth]{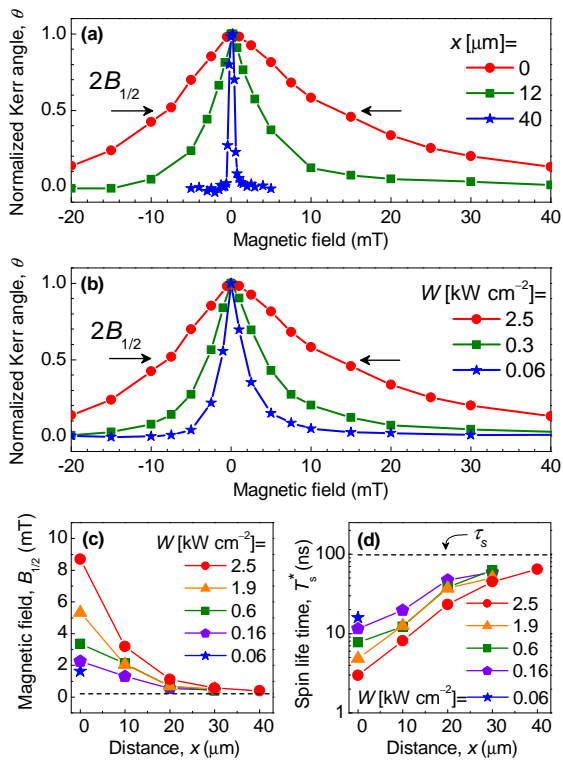}
\caption{(Color online) Spatially resolved Hanle effect obtained
with $W_{probe} = 0.1$~$\mathrm{kW cm^{-2}}$. (a) Local Hanle
curves for different distances $x$. $W_{pump} = 2.5$~$\mathrm{kW
cm^{-2}}$. (b) Local Hanle curves for different pump power densities
$W_{pump}$. $x = 0$~$\mathrm{\mu m}$. (c) Spatial dependencies of
$B_{1/2}$ for different pump power densities. (d) Spatial dependence
of the apparent spin lifetime $T_s^*$ for different pump
power densities. } \label{fig_LocalHanle}
\end{figure}

In order to understand the origin of this effect, we have taken
Hanle curves at various positions in the diffusion profile. These data
are shown in Fig.~\ref{fig_LocalHanle}(a). The small
asymmetry of these curves results from the orientation of the
optically generated spins under an angle $\gamma= 11^{\circ}$ with
respect to the sample normal ($z$-axis). At the generation point $x
= 0$~$\mathrm{\mu m}$ a much higher magnetic field ($2 B_{1/2} =
18$~mT) is required to suppress the spin polarization than in more
distant locations ($2 B_{1/2} = 6$~mT at $x = 12$~$\mathrm{\mu
m}$  and $2 B_{1/2} = 0.8$~mT at $x = 40$~$\mathrm{\mu m}$). This
strong spatial dependence of the spin lifetime actually scales with illumination power density.
 As shown in Fig.~\ref{fig_LocalHanle}(b), when the pump
power density decreases from $W_{pump} = 2.5$~$\mathrm{kW cm^{-2}}$ to
$W_{pump} = 0.06$~$\mathrm{kW cm^{-2}}$ the width of the local
Hanle curves detected at $x = 0$ also reduces from $2 B_{1/2} =
18$~mT to $2 B_{1/2} = 3.2$~mT.

The spatial dependencies of $B_{1/2}$ for different pump power
densities are summarized in Fig.~\ref{fig_LocalHanle}(c).
Remarkably, when the detection point is far away from the
injection point (for $x > 20$~$\mathrm{\mu m}$), $B_{1/2}$ is
nearly independent of the pump power and tends towards $B_{1/2} =
0.25$~mT, as obtained for low pump power densities in
Fig.~\ref{fig_NonDiff}(e), where spin diffusion can
be neglected. %because of the large illumination area.

The experimental results of Fig.~\ref{fig_LocalHanle}(c) can be
interpreted in terms of an apparent spin lifetime $T_s^*$. While
the Hanle curves in the diffusive regime are not described by the
Lorentzian of Eq.~(\ref{Eq_Hanle}) \cite{OO,Dzh1}, also in this
limit the width of the Hanle curves scales with the inverse of the
spin lifetime \cite{SD_theory}. Hence, we evaluate this apparent
spin lifetime using $T_s^* = \hbar / (g_e \mu_B) B_{1/2}^{-1}$.
The  $T_s^*$ values thus obtained are plotted in
Fig.~\ref{fig_LocalHanle}(d) as a function of the distance $x$ for
different pump power densities. These data clearly demonstrate
that illumination induces a spatial variation of the spin
lifetime.

Based on the results of Fig.~\ref{fig_LocalHanle} we propose the
following physical explanation for the increase of spin flow at
low magnetic field presented in Fig.~\ref{fig_2D}(d): In zero
field the optically generated spins $S_z$ diffuse away to a
distance of the order of the spin diffusion length $L_s$,
resulting in a spin gradient $\nabla_{s} \propto S_z / L_s$.
Outside of the injection area, photo-generated holes are absent
and the spin lifetime entering Eq.~(\ref{Eq1}) is equal to the
electron spin relaxation time, $T_s^* \approx \tau_s$. Therefore,
in relatively weak magnetic fields ($B \sim 5$~mT) the net spin
polarization is completely suppressed in this region. Quite the
opposite situation prevails inside the injection area when the
condition of strong spin pumping ($W_{pump} \gg W_0$) is
fulfilled. In this case $T_s^* \ll \tau_s$ and much higher
magnetic fields are required to suppress the net spin
polarization. As a result, the spin gradient increases in weak
magnetic fields according to $\nabla_{s} \sim  S_z / \Delta$. The
enhancement factor can be estimated at $L_s / \Delta$. In
sufficiently strong magnetic fields the spin polarization inside
the injection area is also suppressed ($S_z \rightarrow 0$) and
the spin gradient decreases to zero. For low pump power densities
($W \ll W_0$) according to Eq.~(\ref{Eq1}) $T_s^* \approx \tau_s$
and the spin gradient decreases spatially uniformly with
increasing magnetic field, and the effect disappears.

Summarizing, we used scanning Kerr microscopy to reveal a strong
influence of pump and probe on spin transport at the
power levels which are frequently assumed to satisfy weak
spin pumping conditions. We demonstrate that optical illumination induces spatial
variations of the spin lifetime, which in turn can be used to locally
control the spin flow when combined with a weak magnetic field. We have shown that the effect
can be successfully harnessed to offer a novel possibility to manipulate spin currents.

The authors thank T.~Kiessling for valuable discussions. This
research was supported by the DFG (SPP 1285).

%---------------------------------------------------------------
%***************************************************************
%---------------------------------------------------------------

\end{document}